\documentclass[a4paper,aps,twocolumn,nofootinbib]{revtex4}
\RequirePackage[colorlinks,hyperindex]{hyperref}
\RequirePackage[english]{babel}
\RequirePackage[latin1]{inputenc}
\RequirePackage[T1]{fontenc}
\RequirePackage{mathrsfs}
\RequirePackage{amsmath}
\RequirePackage{amssymb}
\RequirePackage{amsbsy}
\RequirePackage{color}
\RequirePackage{bm}
\hypersetup{colorlinks=true,breaklinks=true,urlcolor=blue,linkcolor=red}
\begin{document}
\title{\bf{Chern-Simons Extension of ESK Theory}}
\author{Luca Fabbri}
\affiliation{DIME, Sez. Metodi e Modelli Matematici, Universit\`{a} di Genova,\\
Via all'Opera Pia 15, 16145 Genova, ITALY}
\date{\today}
\begin{abstract}
The commonly-known Chern-Simons extension of Einstein gravitational theory is written in terms of a square-curvature term added to the linear-curvature Hilbert Lagrangian. In a recent paper, we constructed two Chern-Simons extensions according to whether they consisted of a square-curvature term added to the square-curvature Stelle Lagrangian or of one linear-curvature term added to the linear-curvature Hilbert Lagrangian \cite{Fabbri:2020ezx}. The former extension gives rise to the topological extension of the re-normalizable gravity, the latter extension gives rise to the topological extension of the least-order gravity. This last theory will be written here in its torsional completion. Then a consequence for cosmology and particle physics will be addressed.
\end{abstract}
\maketitle
\section{Introduction}
When investigating the issue of including topological terms in Einstein gravity one is immediately faced with a problem of homogeneity. The Chern-Simons enlargement of Einsteinian gravitation in its original form is that of \cite{Jackiw:2003pm}, and it consists in adding one specific square-curvature contribution to the linear-curvature Hilbert Lagrangian.

Homogeneity can be restored in two ways: either with the inclusion of a square-curvature term into the square-curvature Stelle Lagrangian \cite{Stelle:1977ry, Stelle:1976gc} or with the inclusion of a linear-curvature term into the linear-curvature Hilbert Lagrangian. In \cite{Fabbri:2020ezx} we have discussed both cases finding, in the former case, the topological re-normalizable gravitation, and, in the latter case, the topological least-order derivative gravitation. In the first instance, the modification proposed by Jackiw and Pi was such that the divergence of the gravitational field equations did not lead to the conservation of the energy, but to one constraint that was not verified in general, while we found that when the energy is that of the Dirac spinors, such a constraint will be verified indeed. Not so in the second instance, where the modification we proposed is such that the divergence of the gravitational field equations does not yield energy conservation but a constraint that is not verified, and in this situation it cannot be verified on general grounds.

Of this last, least-order derivative gravitational theory with topological term we will here consider the torsional completion. The torsional completion of gravitation, that is the Sciama-Kibble completion of the Einstein theory, is what we obtain when we allow torsion to be sourced by the spin in the same way in which curvature is sourced by the energy density of the matter field distributions \cite{Hehl:1976kj,Shapiro:2001rz,Hammond:2002rm,Arcos:2005ec}.

Of all possible torsion completions of gravity, here we will take into account the one where torsion is completely antisymmetric so to fit its source given by the completely antisymmetric spin that pertains to the Dirac field \cite{Laemmerzahl:1993zn,Audretsch:1988tu,Fabbri:2006xq,Fabbri:2009se,Fabbri:2008rq,Fabbri:2009yc,Fabbri:2014naa}.

This type of spin-torsion coupling has important effects ranging from cosmology up to particle physics \cite{Fabbri:2017rjf,Fabbri:2017xch,Fabbri:2010rw,Fabbri:2012ag}.

In this paper, the torsion completion of gravity will be performed for the theory that is dynamically defined with the least-order of the derivative, namely the one linear in the curvature or equivalently the one linear in derivatives and squares of the connection, for both the leading term given by the Hilbert Lagrangian and the topological term given by the Chern-Simons type of extension.
\section{Sciama-Kibble Torsion Completion of the Einstein gravity}
We begin by specifying that in the background, metric properties will be defined in terms of the metric $g_{\alpha\nu}\!=\!g_{\nu\alpha}$ and $g^{\alpha\nu}\!=\!g^{\nu\alpha}$ such that $g_{\alpha\nu}g^{\nu\sigma}\!=\!\delta^{\sigma}_{\alpha}$ and which can also be used as a tool for raising/lowering indices in tensorial quantities. Tetrads are defined so that $e_{\alpha}^{r}e_{\nu}^{s}g^{\nu\alpha}\!=\!\eta^{rs}$ and $e^{\alpha}_{r}e^{\nu}_{s}g_{\nu\alpha}\!=\!\eta_{rs}$ where $\eta$ is the Minkowskian matrix, as well as $e^{\alpha}_{r}e_{\alpha}^{s}\!=\!\delta_{r}^{s}$ and $e^{\alpha}_{r}e_{\nu}^{r}\!=\!\delta_{\nu}^{\alpha}$ in such a way that these pairs of dual tetrads allow the passage from coordinate (Greek) indices to Lorentz (Latin) indices. Clifford matrices $\boldsymbol{\gamma}^{a}$ are defined by $\left\{\boldsymbol{\gamma}_{a},\!\boldsymbol{\gamma}_{b}\right\}\!=\!2\eta_{ab}\mathbb{I}$ so that the $\left[\boldsymbol{\gamma}_{a},\!\boldsymbol{\gamma}_{b}\right]\!=\!4\boldsymbol{\sigma}_{ab}$ defines the generators of the complex Lorentz algebra and relationship $2i\boldsymbol{\sigma}_{ab}\!=\!\varepsilon_{abcd}\boldsymbol{\pi}\boldsymbol{\sigma}^{cd}$ implicitly defines $\boldsymbol{\pi}$ (this matrix is usually denoted as gamma matrix with an index five, but in space-time this index has no meaning, and so we employ a notation with no index), and then we have
\begin{eqnarray}
&\boldsymbol{\gamma}_{i}\boldsymbol{\gamma}_{j}\boldsymbol{\gamma}_{k}
\!=\!\boldsymbol{\gamma}_{i}\eta_{jk}-\boldsymbol{\gamma}_{j}\eta_{ik}
\!+\!\boldsymbol{\gamma}_{k}\eta_{ij}
\!+\!i\varepsilon_{ijkq}\boldsymbol{\pi}\boldsymbol{\gamma}^{q}
\end{eqnarray}
with
\begin{eqnarray}
&\{\boldsymbol{\gamma}_{a},\boldsymbol{\sigma}_{bc}\}
=i\varepsilon_{abcd}\boldsymbol{\pi}\boldsymbol{\gamma}^{d}\\
&[\boldsymbol{\gamma}_{a},\boldsymbol{\sigma}_{bc}]
=\eta_{ab}\boldsymbol{\gamma}_{c}\!-\!\eta_{ac}\boldsymbol{\gamma}_{b}\label{commgamma}
\end{eqnarray}
and
\begin{eqnarray}
&\{\boldsymbol{\sigma}_{ab},\boldsymbol{\sigma}_{cd}\}
=\frac{1}{2}[(\eta_{ad}\eta_{bc}\!-\!\eta_{ac}\eta_{bd})\mathbb{I}
\!+\!i\varepsilon_{abcd}\boldsymbol{\pi}]\\
&[\boldsymbol{\sigma}_{ab},\boldsymbol{\sigma}_{cd}]
=\eta_{ad}\boldsymbol{\sigma}_{bc}\!-\!\eta_{ac}\boldsymbol{\sigma}_{bd}
\!+\!\eta_{bc}\boldsymbol{\sigma}_{ad}\!-\!\eta_{bd}\boldsymbol{\sigma}_{ac}\label{commsigma}
\end{eqnarray}
and the $\boldsymbol{\gamma}_{0}$ will be used for the adjoint spinor operation.

This triplication of formalisms may appear to be futile but in fact it is very important, because coordinate tensor fields are defined to be what transforms according to the most general coordinate transformation. Then the tetrad bases convert coordinate tensor fields into Lorentz tensor fields, transforming according to local Lorentz transformations. Finally, the generators of the complex Lorentz algebra can be exponentiated to yield the local complex Lorentz group in terms of which spinor fields transform.

With the concept of scalar product between spinors it is possible to form the bi-linear spinor quantities
\begin{eqnarray}
&\Sigma^{ab}\!=\!2\overline{\psi}\boldsymbol{\sigma}^{ab}\boldsymbol{\pi}\psi\\
&M^{ab}\!=\!2i\overline{\psi}\boldsymbol{\sigma}^{ab}\psi
\end{eqnarray}
\begin{eqnarray}
&S^{a}\!=\!\overline{\psi}\boldsymbol{\gamma}^{a}\boldsymbol{\pi}\psi\\
&U^{a}\!=\!\overline{\psi}\boldsymbol{\gamma}^{a}\psi
\end{eqnarray}
\begin{eqnarray}
&\Theta\!=\!i\overline{\psi}\boldsymbol{\pi}\psi\\
&\Phi\!=\!\overline{\psi}\psi
\end{eqnarray}
such that they are all real tensors, and they verify
\begin{eqnarray}
\nonumber
&\psi\overline{\psi}\!\equiv\!\frac{1}{4}\Phi\mathbb{I}
\!+\!\frac{1}{4}U_{a}\boldsymbol{\gamma}^{a}
\!+\!\frac{i}{8}M_{ab}\boldsymbol{\sigma}^{ab}-\\
&-\frac{1}{8}\Sigma_{ab}\boldsymbol{\sigma}^{ab}\boldsymbol{\pi}
\!-\!\frac{1}{4}S_{a}\boldsymbol{\gamma}^{a}\boldsymbol{\pi}
\!-\!\frac{i}{4}\Theta \boldsymbol{\pi}\label{Fierz}
\end{eqnarray}
and then
\begin{eqnarray}
&2\boldsymbol{\sigma}^{\mu\nu}U_{\mu}S_{\nu}\boldsymbol{\pi}\psi\!+\!U^{2}\psi=0\label{1}\\
&i\Theta S_{\mu}\boldsymbol{\gamma}^{\mu}\psi
\!+\!\Phi S_{\mu}\boldsymbol{\gamma}^{\mu}\boldsymbol{\pi}\psi\!+\!U^{2}\psi=0\label{2}
\end{eqnarray}
and
\begin{eqnarray}
&\Sigma^{ab}\!=\!-\frac{1}{2}\varepsilon^{abij}M_{ij}\label{A}\\
&M^{ab}\!=\!\frac{1}{2}\varepsilon^{abij}\Sigma_{ij}\label{B}
\end{eqnarray}
with
\begin{eqnarray}
&M_{ab}\Phi\!-\!\Sigma_{ab}\Theta\!=\!U^{j}S^{k}\varepsilon_{jkab}\label{A1}\\
&M_{ab}\Theta\!+\!\Sigma_{ab}\Phi\!=\!U_{[a}S_{b]}\label{A2}
\end{eqnarray}
alongside to
\begin{eqnarray}
&M_{ik}U^{i}\!=\!\Theta S_{k}\label{P1}\\
&\Sigma_{ik}U^{i}\!=\!\Phi S_{k}\label{L1}\\
&M_{ik}S^{i}\!=\!\Theta U_{k}\label{P2}\\
&\Sigma_{ik}S^{i}\!=\!\Phi U_{k}\label{L2}
\end{eqnarray}
and with the orthogonality relations
\begin{eqnarray}
&\frac{1}{2}M_{ab}M^{ab}\!=\!-\frac{1}{2}\Sigma_{ab}\Sigma^{ab}\!=\!\Phi^{2}\!-\!\Theta^{2}
\label{norm2}\\
&\frac{1}{2}M_{ab}\Sigma^{ab}\!=\!-2\Theta\Phi
\label{orthogonal2}
\end{eqnarray}
and
\begin{eqnarray}
&U_{a}U^{a}\!=\!-S_{a}S^{a}\!=\!\Theta^{2}\!+\!\Phi^{2}\label{norm1}\\
&U_{a}S^{a}\!=\!0\label{orthogonal1}
\end{eqnarray}
as geometric identities called Fierz identities. Notice that while defining six bi-linear spinor quantities maintains a certain symmetry in the expression of the Fierz identities, (\ref{A}, \ref{B}) show that of the two antisymmetric tensors only one is actually needed. In the general case where at least one between $\Theta$ and $\Phi$ is not identically zero (\ref{A1}, \ref{A2}) can be combined to give the expressions
\begin{eqnarray}
&\Sigma^{ab}\!=\!2\phi^{2}(\cos{\beta}u^{[a}s^{b]}\!-\!\sin{\beta}u_{j}s_{k}\varepsilon^{jkab})\\
&M^{ab}\!=\!2\phi^{2}(\cos{\beta}u_{j}s_{k}\varepsilon^{jkab}\!+\!\sin{\beta}u^{[a}s^{b]})
\end{eqnarray}
where we have introduced
\begin{eqnarray}
&S^{a}\!=\!2\phi^{2}s^{a}\\
&U^{a}\!=\!2\phi^{2}u^{a}
\end{eqnarray}
and
\begin{eqnarray}
&\Theta\!=\!2\phi^{2}\sin{\beta}\\
&\Phi\!=\!2\phi^{2}\cos{\beta}
\end{eqnarray}
for simplicity, and showing that in this case no antisymmetric tensor at all is needed. With these expressions all Fierz identities (\ref{P1}-\ref{L2}) and (\ref{norm2}, \ref{orthogonal2}) reduce to be trivial while (\ref{norm1}, \ref{orthogonal1}) reduce to 
\begin{eqnarray}
&u_{a}u^{a}\!=\!-s_{a}s^{a}\!=\!1\\
&u_{a}s^{a}\!=\!0
\end{eqnarray}
so that $u^{a}$ is time-like and $s^{a}$ is space-like in such a case.

This is important because this means that one can always find some combination of up to three Lorentz boosts bringing $u^{a}$ to have the temporal component only. This tells us that $u^{a}$ is the velocity vector and that a rest frame is possible. However, these Fierz identities also tell that $s^{a}$ is the Pauli-Lubanski axial-vector, and in general it is always possible to find some combinations of up to two Lorentz rotations bringing it to have only the third of its components. A
consequence of this is that $s^{a}$ is the spin axial-vector and one can always find a frame where it is aligned with the third of the axes. In this situation, the last rotation is the one around the third axis. As such it can be used to act onto the global phase of the spinorial field. Putting all together allows one to see that general spinors can always be written according to
\begin{eqnarray}
&\!\psi\!=\!\phi e^{-\frac{i}{2}\beta\boldsymbol{\pi}}
\boldsymbol{S}\left(\!\begin{tabular}{c}
$1$\\
$0$\\
$1$\\
$0$
\end{tabular}\!\right)
\label{spinor}
\end{eqnarray}
in chiral representation, for some local complex Lorentz transformation $\boldsymbol{S}$, with the $\beta$ and $\phi$ being a pseudo-scalar and scalar fields called Yvon-Takabayashi angle and module, and where the spinor is said in polar form. In such a form, the initial $8$ real components are re-arranged into the configuration in which the $2$ true degrees of freedom given by the YT angle and module remain isolated from the $6$ components that can always be transferred into the frame given by the velocity and spin, or equivalently by the rapidities and angles that are encoded in the Lorentz transformation. Notice that the YT angle has zero mass-dimension and so the module inherits the full $3/2$ mass-dimension that characterizes spinors. The reader who is interested in the details of the construction of this polar decomposition can have a look at references \cite{L,Cavalcanti:2014wia,HoffdaSilva:2017waf,daSilva:2012wp,Ablamowicz:2014rpa,Rodrigues:2005yz,HoffdaSilva:2009is,daRocha:2008we,daRocha:2013qhu,Fabbri:2016msm}.

The most general connection shall be indicated by $\Gamma^{\sigma}_{\alpha\nu}$ in general. Then the corresponding most general covariant derivative will be assumed to satisfy $D_{\nu}g_{\alpha\sigma}\!=\!0$ known as metric-compatibility condition. Notice that the antisymmetric part $\Gamma^{\sigma}_{\alpha\nu}\!-\!\Gamma^{\sigma}_{\nu\alpha}\!=\!Q^{\sigma}_{\phantom{\sigma}\alpha\nu}$ is in general not equal to zero and it turns out to be a tensor, and this is what is normally called torsion tensor. We will assume torsion to be completely antisymmetric, and in such a case torsion can  always be written like $6Q_{\sigma\alpha\nu}\!=\! -\varepsilon_{\sigma\alpha\nu\pi}W^{\pi}$ as Hodge dual of an axial-vector. The two assumptions encoded in the condition of metric-compatibility and in the complete antisymmetry of torsion are enforced so that the general connection can be expressed according to the following
\begin{eqnarray}
&\Gamma^{\sigma}_{\alpha\nu}\!=\!\frac{1}{2}g^{\sigma\rho}[(\partial_{\alpha}g_{\nu\rho}
\!+\!\partial_{\nu}g_{\alpha\rho}\!-\!\partial_{\rho}g_{\alpha\nu})
\!+\!\frac{1}{6}W^{\pi}\varepsilon_{\pi\rho\alpha\nu}]
\end{eqnarray}
where the part that is entirely written in terms of the partial derivatives of the metric is the symmetric connection whose associated covariant derivative satisfies conditions $\nabla_{\nu}g_{\alpha\sigma}\!\equiv\!0$ identically. The passage to tetradic formalism is possible in terms of the spin connection $\Omega^{a}_{\phantom{a}c\pi}$ for which the corresponding covariant derivative is assumed to satisfy $D_{\nu}e_{\alpha}^{c}\!=\!0$ similarly to what was done before. Because of the different type of indices no antisymmetric part can be defined for the spin connection. But the vanishing of the covariant derivative of the tetrads can be written as
\begin{eqnarray}
&\Omega^{a}_{\phantom{a}b\pi}\!=\!e^{\nu}_{b}
e^{a}_{\sigma}(\partial_{\pi}e^{\sigma}_{i}e_{\nu}^{i}\!+\!\Gamma^{\sigma}_{\nu\pi})
\end{eqnarray}
linking the connection to the spin connection, so that it is still possible to decompose the spin connection into the sum of a torsion term plus the torsionless spin connection, that is the one whose associated covariant derivative still satisfies $\nabla_{\nu}e_{\alpha}^{c}\!\equiv\!0$ identically as above. Finally, spinorial connections can also be defined as $\boldsymbol{\Omega}_{\mu}$ with corresponding covariant derivative assumed to satisfy $\boldsymbol{D}_{\mu}\boldsymbol{\gamma}_{a}\!=\!0$ also in analogy to what done before. This condition can then be worked out to give the following expression
\begin{eqnarray}
&\boldsymbol{\Omega}_{\mu}\!=\!iqA_{\mu}\boldsymbol{\mathbb{I}}
\!+\!\frac{1}{2}\Omega^{ab}_{\phantom{ab}\mu}\boldsymbol{\sigma}_{ab}
\label{spinorialconnection}
\end{eqnarray}
as the decomposition in terms of the spin connection with an additional vector field, and again we could decompose the spinorial connection into the sum of a torsional term plus the torsionless spinorial connection, with associated covariant derivative $\boldsymbol{\nabla}_{\mu}\boldsymbol{\gamma}_{a}\!\equiv\!0$ identically as above.

When the polar form is taken into the spinorial derivative and considering that we can formally write
\begin{eqnarray}
&\boldsymbol{S}\partial_{\mu}\boldsymbol{S}^{-1}\!=\!i\partial_{\mu}\alpha\mathbb{I}
\!+\!\frac{1}{2}\partial_{\mu}\theta_{ij}\boldsymbol{\sigma}^{ij}\label{spintrans}
\end{eqnarray}
we can define
\begin{eqnarray}
&\partial_{\mu}\alpha\!-\!qA_{\mu}\!\equiv\!P_{\mu}\label{P}\\
&\partial_{\mu}\theta_{ij}\!-\!\Omega_{ij\mu}\!\equiv\!R_{ij\mu}\label{R}
\end{eqnarray}
needed to write
\begin{eqnarray}
&\!\!\!\!\!\!\!\!\boldsymbol{D}_{\mu}\psi\!=\!(-\frac{i}{2}D_{\mu}\beta\boldsymbol{\pi}
\!+\!D_{\mu}\ln{\phi}\mathbb{I}
\!-\!iP_{\mu}\mathbb{I}\!-\!\frac{1}{2}R_{ij\mu}\boldsymbol{\sigma}^{ij})\psi
\label{decspinder}
\end{eqnarray}
as spinorial covariant derivative, from which
\begin{eqnarray}
&D_{\mu}s_{i}\!=\!R_{ji\mu}s^{j}\label{ds}\\
&D_{\mu}u_{i}\!=\!R_{ji\mu}u^{j}\label{du}
\end{eqnarray}
valid as general geometric identities. Notice that despite (\ref{P}, \ref{R}) contain the same information of connection and gauge potential nonetheless they are proven to be tensors and invariant under the gauge transformation. As we said above, writing spinor fields in polar form allows to make a re-configuration in which the degrees of freedom remain isolated from the components transferable into the gauge and the frames, and now we see that the components that are transferred into gauge and frames combine with gauge potential and spin connection, with no alteration to their information content while rendering the resulting objects true tensors invariant under gauge transformations. Thus they have been called gauge-invariant vector momentum and tensorial connection. To have more details see \cite{Fabbri:2018crr}.

Then the commutator of spinorial covariant derivatives justifies the definitions
\begin{eqnarray}
&R^{i}_{\phantom{i}j\mu\nu}\!=\!\partial_{\mu}\Omega^{i}_{\phantom{i}j\nu}
\!-\!\partial_{\nu}\Omega^{i}_{\phantom{i}j\mu}
\!+\!\Omega^{i}_{\phantom{i}k\mu}\Omega^{k}_{\phantom{k}j\nu}
\!-\!\Omega^{i}_{\phantom{i}k\nu}\Omega^{k}_{\phantom{k}j\mu}\\
&F_{\mu\nu}\!=\!\partial_{\mu}A_{\nu}\!-\!\partial_{\nu}A_{\mu}
\end{eqnarray}
that is the Riemann curvature and the Maxwell strength.

Taking the commutator with polar variables gives
\begin{eqnarray}
\!\!\!\!&qF_{\mu\nu}\!=\!-(D_{\mu}P_{\nu}\!-\!D_{\nu}P_{\mu}
\!-\!P_{\alpha}Q^{\alpha}_{\phantom{\alpha}\mu\nu})\label{Maxwell}\\
\nonumber
&\!\!\!\!\!\!\!\!R^{i}_{\phantom{i}j\mu\nu}\!=\!-(D_{\mu}R^{i}_{\phantom{i}j\nu}
\!-\!\!D_{\nu}R^{i}_{\phantom{i}j\mu}+\\
&+R^{i}_{\phantom{i}k\mu}R^{k}_{\phantom{k}j\nu}
\!-\!R^{i}_{\phantom{i}k\nu}R^{k}_{\phantom{k}j\mu}
\!-\!R^{i}_{\phantom{i}j\alpha}Q^{\alpha}_{\phantom{\alpha}\mu\nu})\label{Riemann}
\end{eqnarray}
in terms of the Maxwell strength and Riemann curvature, so that they encode electrodynamic and gravitational information filtering out gauge and frame information. But on the other hand, the gauge-invariant vector momentum and tensorial connection (\ref{P}, \ref{R}) contain the information about electrodynamics and gravity and also about gauge and frames. Because there exist non-zero gauge-invariant vectorial momentum and tensorial connection which still give zero Maxwell strength and Riemann curvature, then there exists a type of information that is related to gauge and frames solely but which is also fully covariant.

The Dirac spinor dynamics is given by the Dirac spinor field equations
\begin{eqnarray}
&i\boldsymbol{\gamma}^{\mu}\boldsymbol{D}_{\mu}\psi\!-\!m\psi\!=\!0
\label{D}
\end{eqnarray}
and by multiplying it by $\boldsymbol{\gamma}^{a}$ and $\boldsymbol{\gamma}^{a}\boldsymbol{\pi}$ and by $\overline{\psi}$ splitting real and imaginary parts gives
\begin{eqnarray}
&\!\!\!\!D_{\alpha}\Phi
\!-\!2(\overline{\psi}\boldsymbol{\sigma}_{\mu\alpha}\boldsymbol{D}^{\mu}\psi
\!-\!\boldsymbol{D}^{\mu}\overline{\psi}\boldsymbol{\sigma}_{\mu\alpha}\psi)\!=\!0\\
&\!\!\!\!D_{\nu}\Theta\!-\!
2i(\overline{\psi}\boldsymbol{\sigma}_{\mu\nu}\boldsymbol{\pi}\boldsymbol{D}^{\mu}\psi\!-\!
\boldsymbol{D}^{\mu}\overline{\psi}\boldsymbol{\sigma}_{\mu\nu}\boldsymbol{\pi}\psi)
\!+\!2mS_{\nu}\!=\!0
\end{eqnarray}
which are called Gordon decompositions.

In them it is possible to plug the polar form getting
\begin{eqnarray}
&\!\!\!\!B_{\mu}\!-\!2P^{\iota}u_{[\iota}s_{\mu]}\!+\!D_{\mu}\beta
\!+\!2s_{\mu}m\cos{\beta}\!=\!0\label{dep1}\\
&\!\!\!\!R_{\mu}\!-\!2P^{\rho}u^{\nu}s^{\alpha}\varepsilon_{\mu\rho\nu\alpha}\!+\!2s_{\mu}m\sin{\beta}
\!+\!D_{\mu}\ln{\phi^{2}}\!=\!0\label{dep2}
\end{eqnarray}
with $R_{\mu a}^{\phantom{\mu a}a}\!=\!R_{\mu}$ and $\frac{1}{2}\varepsilon_{\mu\alpha\nu\iota}R^{\alpha\nu\iota}\!=\!B_{\mu}$ and which could be proven to be equivalent to the polar form of the Dirac spinorial field equations. Notice that the Dirac spinorial field equations are $8$ real equations and that is as many as the $2$ vectorial equations given by the (\ref{dep1}, \ref{dep2}) above, thus specifying all space-time derivatives of the two degrees of freedom given by YT angle and module. The reader who is interested in details may have a look at reference \cite{Fabbri:2016laz}.

An example of gauge-invariant vector momentum and tensorial connection that are non-zero but which have a Maxwell strength and a Riemann curvature null is in \cite{Fabbri:2019kfr}.

For general summaries, readers might look at \cite{Fabbri:2017pwp,Fabbri:2020ypd}.
\section{Chern-Simons Topological Extension}
The Chern-Simons topological extension of the Einstein gravity is done by adding to the Hilbert Lagrangian
\begin{eqnarray}
&\mathscr{L}\!=\!R^{i}_{\phantom{i}j\mu\nu}e^{\mu}_{i}e^{\nu}_{k}\eta^{jk}
\end{eqnarray}
a term of the form
\begin{eqnarray}
&\mathscr{L}\!=\!kbD_{\mu}K^{\mu}
\end{eqnarray}
with $b$ pseudo-scalar and $K^{\mu}$ axial-vector. Homogeneity requires this term to have $2$ mass-dimension, and the way we have to do this is to take $b$ as a pseudo-scalar of zero mass dimension and $K^{\mu}$ an axial-vector of unitary mass-dimension. The polar form of spinor fields shows that we do have such objects, the Yvon-Takabayashi angle $\beta$ and the dual of the completely antisymmetric component of the tensorial connection $B^{\mu}$ above. So we may write
\begin{eqnarray}
&\mathscr{L}\!=\!k\beta D_{\mu}B^{\mu}
\end{eqnarray}
for the topological term. Altogether we have 
\begin{eqnarray}
&\mathscr{L}\!=\!R\!+\!k\beta D_{\mu}B^{\mu}
\end{eqnarray}
which is indeed homogeneous, and it will be taken as the Lagrangian for torsion-gravity. Then, we end up having
\begin{eqnarray}
&\mathscr{L}\!=\!R\!+\!k\beta D_{\mu}B^{\mu}
\!-\!i\overline{\psi}\boldsymbol{\gamma}^{\mu}\boldsymbol{D}_{\mu}\psi
\!+\!m\overline{\psi}\psi
\end{eqnarray}
as the Lagrangian of torsion-gravity for a space-time filled with spinorial matter. For more details, see \cite{Fabbri:2020ezx}.

In polar form it becomes
\begin{eqnarray}
\nonumber
&\mathscr{L}\!=\!R\!+\!k\beta D_{\mu}B^{\mu}\!-\!\phi^{2}[s^{\mu}(D_{\mu}\beta\!+\!B_{\mu})+\\
&+2u^{\mu}P_{\mu}\!-\!2m\cos{\beta}]
\end{eqnarray}
as it can be checked straightforwardly.

Its variation with respect to tetrads and torsion gives the gravitational and torsional field equations
\begin{eqnarray}
\nonumber
&R^{\nu\sigma}\!-\!\frac{1}{2}g^{\nu\sigma}R
\!-\!\frac{k}{2}[D_{\mu}(\beta B^{\mu})g^{\nu\sigma}
\!-\!\varepsilon^{\mu\nu\alpha\eta}D_{\mu}\beta R^{\sigma}_{\phantom{\sigma}\alpha\eta}]=\\
&=\frac{i}{4}(\overline{\psi}\boldsymbol{\gamma}^{\nu}\!\boldsymbol{D}^{\sigma}\psi
\!-\!\!\boldsymbol{D}^{\sigma}\overline{\psi}\boldsymbol{\gamma}^{\nu}\psi)
\label{e}\\
&Q^{\alpha\nu\mu}\!+\!\frac{k}{2}\varepsilon^{\alpha\nu\mu\sigma}D_{\sigma}\beta
\!=\!-\frac{i}{4}\overline{\psi}\{\boldsymbol{\sigma}^{\alpha\nu},\boldsymbol{\gamma}^{\mu}\}\psi
\label{sk}
\end{eqnarray}
and with respect to the spinor gives
\begin{eqnarray}
&\!\!\!\!i\boldsymbol{\gamma}^{\mu}\boldsymbol{D}_{\mu}\psi
\!-\!\frac{k}{2}D\!\cdot\!B\phi^{-2}e^{i\boldsymbol{\pi}\beta}i\boldsymbol{\pi}\psi
\!-\!m\psi\!=\!0\label{d}
\end{eqnarray}
as the full set of geometry and matter field equations.

In polar form we have that
\begin{eqnarray}
\nonumber
&R^{\nu\sigma}\!-\!\frac{1}{2}g^{\nu\sigma}R
\!-\!\frac{k}{2}[D_{\mu}(\beta B^{\mu})g^{\nu\sigma}
\!-\!\varepsilon^{\mu\nu\alpha\eta}D_{\mu}\beta R^{\sigma}_{\phantom{\sigma}\alpha\eta}]=\\
&=\phi^{2}(u^{\nu}P^{\sigma}\!+\!\frac{1}{2}s^{\nu}D^{\sigma}\beta
\!-\!\frac{1}{4}R_{ij}^{\phantom{ij}\sigma}\varepsilon^{\nu ijk}s_{k})
\label{e1}\\
&\frac{1}{3}W_{\sigma}\!-\!kD_{\sigma}\beta\!=\!\phi^{2}s_{\sigma}
\label{sk1}
\end{eqnarray}
and
\begin{eqnarray}
&-2P^{\iota}u_{[\iota}s_{\mu]}\!+\!B_{\mu}\!+\!D_{\mu}\beta
\!+\!2s_{\mu}m\cos{\beta}\!=\!0\label{d1}\\
\nonumber
&-2P^{\rho}u^{\nu}s^{\alpha}\varepsilon_{\mu\rho\nu\alpha}\!+\!R_{\mu}
\!-\!s_{\mu}kD\!\cdot\!B\phi^{-2}+\\
&+D_{\mu}\ln{\phi^{2}}\!+\!2s_{\mu}m\sin{\beta}\!=\!0\label{d2}
\end{eqnarray}
again as it is quite direct to verify.

Taking the divergence of (\ref{e}) and (\ref{sk}) and employing the Jacobi-Bianchi cyclic identities gives
\begin{eqnarray}
\nonumber
&\!\!D_{\alpha}[\varepsilon^{\sigma\alpha\rho\nu}R^{\mu}_{\phantom{\mu}\rho\nu}D_{\sigma}\beta
\!-\!g^{\alpha\mu}D_{\nu}(\beta B^{\nu})]-\\
\nonumber
&-Q^{\eta\rho\mu}\varepsilon_{\sigma\rho\pi\nu}D^{\sigma}\beta R_{\eta}^{\phantom{\eta}\pi\nu}+\\
&+\frac{1}{2}R^{\rho\pi\eta\mu}\varepsilon_{\rho\pi\eta\alpha}D^{\alpha}\beta
\!=\!-D^{\mu}\beta D_{\nu}B^{\nu}\label{ce}\\
&\!\!\!\!D_{\sigma}\beta (\varepsilon^{\sigma\alpha\rho\nu}R^{\eta}_{\phantom{\eta}\rho\nu}
\!-\!\varepsilon^{\sigma\eta\rho\nu}R^{\alpha}_{\phantom{\alpha}\rho\nu})
\!=\!\frac{1}{2}\varepsilon^{\alpha\eta\mu\sigma}Q_{\mu\sigma\rho}D^{\rho}\beta\label{csk}
\end{eqnarray}
where also (\ref{d}) have been thoroughly used.

These are a pair of restrictions over the derivatives of the Yvon-Takabayashi angle and tensorial connection.

Notice that with no tensorial connection we would obtain the theory that has been presented in \cite{Lattanzi:2009mg,Castillo-Felisola:2015ema,Fabbri:2020drl}.
\section{Torsionless Discontinuity}
We have presented the general ESK theory for a space-time filled with Dirac spinor matter fields with topological extension of Chern-Simons type. And that is, we have given the topological extension of a spinor interaction in presence of gravity with torsion. Or then again, we have considered the torsional extension of \cite{Fabbri:2020ezx}. Therefore, it is natural to assume that in the torsionless limit we should obtain the results that were obtained in reference \cite{Fabbri:2020ezx}.

Nonetheless, this is generally not possible because here the presence of torsion as an independent degree of freedom makes one more way of varying the Lagrangian, and ultimately one more field equation, namely (\ref{sk1}), and one more constraint, namely (\ref{csk}). Additional field equations and constraints are not necessarily going to be identically satisfied in the torsionless limit, creating a condition that can be called torsionless discontinuity \cite{Fabbri:2014dxa}. In this case, the limit of torsionlessness does \emph{not} at all coincide with the theory one would have had in total absence of torsion.

Consequently, an interesting question is now: what are the effects of the torsionless limit in this case?

To answer this question, we will take the limit $W_{\sigma}\!\rightarrow\!0$ in all field equations and constraints. To do that, first we replace all the covariant derivatives with their torsionless covariant derivatives and specify that all tensorial connections will be intended to be torsionless and indicated as $\widetilde{R}_{ij\mu}$ in the following. Then we have that (\ref{sk1}) gives
\begin{eqnarray}
k\nabla_{\sigma}\beta\!=\!-\phi^{2}s_{\sigma}\label{supp}
\end{eqnarray}
locking the YT angle to the spin density and 
\begin{eqnarray}
\nonumber
&\widetilde{R}^{\nu\sigma}\!-\!\frac{1}{2}g^{\nu\sigma}\widetilde{R}
\!+\!\frac{k}{2}[-\nabla_{\mu}(\beta \widetilde{B}^{\mu})g^{\nu\sigma}+\\
\nonumber
&+\frac{1}{2}\nabla_{\mu}\beta(\varepsilon^{\mu\sigma\alpha\eta}
\widetilde{R}^{\nu}_{\phantom{\nu}\alpha\eta}
\!+\!\varepsilon^{\mu\nu\alpha\eta}\widetilde{R}^{\sigma}_{\phantom{\sigma}\alpha\eta})]=\\
\nonumber
&=\frac{1}{2}\phi^{2}(u^{\nu}P^{\sigma}\!+\!u^{\sigma}P^{\nu}
\!+\!\frac{1}{2}s^{\nu}\nabla^{\sigma}\beta\!+\!\frac{1}{2}s^{\sigma}\nabla^{\nu}\beta-\\
&-\frac{1}{4}\widetilde{R}_{ij}^{\phantom{ij}\sigma}\varepsilon^{\nu ijk}s_{k}
\!-\!\frac{1}{4}\widetilde{R}_{ij}^{\phantom{ij}\nu}\varepsilon^{\sigma ijk}s_{k})
\end{eqnarray}
which have also been symmetrized using the constraints.

In fact the constraints read
\begin{eqnarray}
\nonumber
&\frac{1}{2}\nabla_{\alpha}[(\varepsilon^{\sigma\alpha\rho\nu}
\widetilde{R}^{\mu}_{\phantom{\mu}\rho\nu}
\!+\!\varepsilon^{\sigma\mu\rho\nu}
\widetilde{R}^{\alpha}_{\phantom{\alpha}\rho\nu})\nabla_{\sigma}\beta-\\
&-2g^{\mu\alpha}\nabla_{\nu}(\beta \widetilde{B}^{\nu})]
\!=\!-\nabla^{\mu}\beta \nabla_{\nu}\widetilde{B}^{\nu}\\
&\nabla_{\sigma}\beta (\varepsilon^{\sigma\alpha\rho\nu}
\widetilde{R}^{\eta}_{\phantom{\eta}\rho\nu}
\!-\!\varepsilon^{\sigma\eta\rho\nu}
\widetilde{R}^{\alpha}_{\phantom{\alpha}\rho\nu})\!=\!0\label{add}
\end{eqnarray}
as it can be easily checked.

Notice that these are exactly all the field equations and constraints of \cite{Fabbri:2020ezx} plus the additional field equation (\ref{supp}) and constraint (\ref{add}). So the limit of torsionlessness does not give the theory in absence of torsion, and torsionless discontinuity arises in the topological torsion-gravity.

Notice however that if we were to take also the spinless limit $s_{\sigma}\!\rightarrow\!0$ then we will see further simplifications.

In fact in this case (\ref{supp}) would give $\nabla_{\nu}\beta\!=\!0$ so that (\ref{add}) would be identically verified, hence continuity recovered.

However, we would also have that the other constraint would have the only solution $\nabla_{\mu}\nabla\!\cdot\!\widetilde{B}\!=\!0$ so that
\begin{eqnarray}
&\widetilde{R}^{\nu\sigma}\!-\!\frac{1}{2}g^{\nu\sigma}\widetilde{R}\!-\!\Lambda g^{\nu\sigma}
\!=\!\frac{1}{2}\phi^{2}(u^{\nu}P^{\sigma}\!+\!u^{\sigma}P^{\nu})
\end{eqnarray}
where $2\Lambda\!=\!k\beta\nabla\!\cdot\!\widetilde{B} $ was defined. Because $\beta$ and $\nabla\!\cdot\!\widetilde{B}$ are constant, then these equations are just the usual gravitational field equations but with an effective cosmological constant that is generated by the underlying topology.

This is in fact the macroscopic limit of the topological extension of Einstein gravity with torsion in interactions with Dirac spinorial matter fields as discussed in \cite{Fabbri:2020ezx}.
\section{Nonlocality}
In the last two sections we have witnessed that in taking the divergence of the field equations to get conservation laws we instead obtained constraints, as is expected from the fact that the tensorial connection does not verify any system of field equations. And that its presence does provide the conditions to generate an effective cosmological constant, as expected for fields that do not vanish at infinity. So, we must also expect a nonlocal behaviour.

Consider for example the situation of a spinor that displays a flipping of the spin $s_{a}$ in general. The process can be described by a rotation of angle $\theta\!=\!\omega t$ around the first axis. The standard situation with no flip of the spin is obtained when $s_{a}$ points toward a fixed direction. In this instance, we can still use the above solutions but taking $\omega\!=\!0$ in general. We may call $R'_{ij\mu}$ the tensorial connection in the former case and $R_{ij\mu}$ the tensorial connection in the latter case. Because of (\ref{ds}) it is clear that in both cases we will have $R'_{ij\mu}\!=\!R_{ij\mu}\!+\!\Delta R_{ij\mu}$ where $\Delta R_{ij\mu}$ has $\Delta R_{23t}\!=\!\omega$ with all the other components being equal to zero identically. Suppose now that a measurement process intervenes so that the former solution reduces to the latter solution. Such a process clearly would require that $\omega\!\rightarrow\!0$ as the reduction of the solutions. So we would also have $R'_{ij\mu}\!\rightarrow\!R_{ij\mu}$ as the reduction for the corresponding tensorial connections. This would hold for one particle.

If then the above solution is meant to describe a state of two particles with opposite spin, then the reduction of one spin would imply the reduction of the other spin for spin conservation. As discussed above, the transmission of the reduction from one spin to the other will be done in terms of the tensorial connection. Such a transmission does not need to be causal because tensorial connections do not have any genuine form of dynamical propagation.

It is however important to specify that any mechanism taking place via the tensorial connection need not respect causality because the tensorial connection is not governed by field equations, and not because causality is violated in the dynamics. In fact, causality is still ensured for fields that propagate, and more in general the causal structure is granted for fields that are solutions of field equations.

The possibility of having nonlocal effects may be given without violating the causal propagation because we are describing them with objects that have no dynamics.
\section{Conclusion}
In this paper, we have constructed the Sciama-Kibble torsional completion of the Einsteinian gravitational theory that is obtained by extending with the Chern-Simons topological terms the Hilbert Lagrangian. The result was a theory in which the topological term was formed by a purely spinorial degree of freedom, the Yvon-Takabayashi angle, and the divergence of a purely geometric degree of freedom, the dual of the completely antisymmetric part of the tensorial connection that also contained the torsion axial-vector. This term therefore is the product of a pair of dynamically coupled variables and as such it accounts for an interaction between the two. As a consequence, a check on the conservation laws failed to give the conservation of the energy, and it yielded some constraints that were not identically verified in general circumstances.

We found that in the limit of torsionlessness the theory had discontinuities, since the constraints were still not in general satisfied. It was only by taking the spinless limit that continuity would be ensured by the total lack of any constraint. In this case, we found that a term giving an effective cosmological constant was eventually generated in the macroscopic limit that ensured the continuity.

The lack of dynamical field equations for the tensorial connection, together with its non-trivial structure at infinity, and its nonlocal properties, all point toward the fact that the tensorial connection is rather unique in the panorama of physical theories developed in recent times.

All this appears to indicate that the tensorial connection has the character of an object that cannot describe physical degrees of freedom, and yet it is found to be not equal to zero in quite general circumstances.


\begin{thebibliography}{40}
\bibitem{Jackiw:2003pm}
R.Jackiw, S.Y.Pi, ``Chern-Simons modification of general relativity'', \textit{Phys.Rev.D}\textbf{68}, 104012 (2003).
\bibitem{Stelle:1977ry} 
K.S.Stelle, ``Classical Gravity with Higher Derivatives'',\\
\textit{Gen.Rel.Grav.} \textbf{9}, 353 (1978).
\bibitem{Stelle:1976gc}
K.S.Stelle, ``Renormalization of Higher Derivative Quantum Gravity'', \textit{Phys.Rev.D} \textbf{16}, 953 (1977).
\bibitem{Fabbri:2020ezx}
L.Fabbri, ``The most complete mass-dimension four topological gravity'', 
\textit{Gen.Rel.Grav}\textbf{52}, 96 (2020).
\bibitem{Hehl:1976kj}
F.W.Hehl, P.Von Der Heyde, G.D.Kerlick, J.M.Nester,\\
``General Relativity with Spin and Torsion: Foundations\\
and Prospects'', \textit{Rev.Mod.Phys.}\textbf{48}, 393 (1976).
\bibitem{Shapiro:2001rz}
I.L.Shapiro, ``Physical aspects of the space-time torsion'',\\
\textit{Phys.Rept.}\textbf{357}, 113 (2002).
\bibitem{Hammond:2002rm}
R.T.Hammond, ``Torsion gravity'',\\
\textit{Rept.Prog.Phys.}\textbf{65}, 599 (2002).
\bibitem{Arcos:2005ec}
H.I.Arcos, J.G.Pereira, ``Torsion gravity:\\ 
A Reappraisal'', \textit{Int.J.Mod.Phys.D}\textbf{13}, 2193 (2004).
\bibitem{Laemmerzahl:1993zn}
C.Laemmerzahl, A.Macias, ``On the dimensionality of\\ 
space-time'', \textit{J. Math. Phys.} \textbf{34}, 4540 (1993).
\bibitem{Audretsch:1988tu}
J.Audretsch, C.Lammerzahl, ``Constructive\\ Axiomatic Approach To Space-time Torsion'',\\ \textit{Class. Quant. Grav.} \textbf{5}, 1285 (1988).
\bibitem{Fabbri:2006xq} 
L.Fabbri, ``On a completely antisymmetric Cartan torsion\\ 
tensor'', \textit{In Annales de la Fondation de Broglie,\\ Special Issue on Torsion (2007)}.
\bibitem{Fabbri:2009se}
L.Fabbri, ``On the Principle of Equivalence'',\\ 
\textit{In Contemporary Fundamental Physics,\\ Einstein and Hilbert: Dark Matter (2012)}
\bibitem{Fabbri:2008rq} 
L.Fabbri, ``On the problem of Unicity in Einstein-Sciama-Kibble Theory'', \textit{Annales Fond. Broglie}\textbf{33}, 365 (2008).
\bibitem{Fabbri:2009yc} 
L.Fabbri, ``On the consistency of Constraints in Matter Field Theories'', \textit{Int.J.Theor.Phys.}\textbf{51}, 954 (2012).
\bibitem{Fabbri:2014naa} 
L.Fabbri, ``Least-order torsion-gravity for fermion fields,\\ and the nonlinear potentials in the standard models'',\\ \textit{Int.J.Geom.Meth.Mod.Phys.}\textbf{11}, 1450073 (2014).
\bibitem{Fabbri:2017rjf}
L.Fabbri, ``Singularity-free spinors in gravity with propagating torsion'',
\textit{Mod.Phys.Lett.A}\textbf{32}, 1750221 (2017).
\bibitem{Fabbri:2017xch}
L.Fabbri, ``A geometrical assessment of spinorial energy conditions'', \textit{Eur.Phys.J.Plus}\textbf{132}, 156 (2017).
\bibitem{Fabbri:2010rw}
L.Fabbri, ``On geometric relativistic foundations of\\ matter field equations and plane wave solutions'', \\ \textit{Mod.Phys.Lett.A}\textbf{27}, 1250028 (2012).
\bibitem{Fabbri:2012ag}
L.Fabbri, ``On a purely geometric approach to the\\ Dirac matter field and its quantum properties'',\\ \textit{Int.J.Theor.Phys.}\textbf{53}, 1896 (2014).
\bibitem{L}
P.Lounesto, \textit{Clifford Algebras and Spinors} (2001).
\bibitem{Cavalcanti:2014wia}
R.T.Cavalcanti, ``Classification of Singular Spinor\\ Fields and Other Mass Dimension One Fermions'',\\ \textit{Int.J.Mod.Phys.D}\textbf{23}, 1444002 (2014).
\bibitem{HoffdaSilva:2017waf}
J.M.Hoff da Silva, R.T.Cavalcanti, ``Revealing how\\ different spinors can be: the Lounesto 
spinor\\ classification'', \textit{Mod.Phys.Lett.A}\textbf{32}, 1730032 (2017).
\bibitem{daSilva:2012wp}
J.M.Hoff da Silva, R.da Rocha, ``Unfolding Physics\\ from the Algebraic Classification of Spinor\\ Fields'', \textit{Phys. Lett. B}\textbf{718}, 1519 (2013).
\bibitem{Ablamowicz:2014rpa}
R.Ab{\l}amowicz, I.Gon{\c c}alves, R.da Rocha, ``Bilinear\\ Covariants and Spinor Fields Duality in Quantum\\ Clifford Algebras'', \textit{J. Math. Phys.}\textbf{55}, 103501 (2014).
\bibitem{Rodrigues:2005yz}
W.A.Rodrigues, R.da Rocha, J.Vaz, ``Hidden\\ consequence of active local Lorentz invariance'',\\ \textit{Int.J.Geom.Meth.Mod.Phys.}\textbf{2}, 305 (2005).
\bibitem{HoffdaSilva:2009is} 
J.M.Hoff da Silva, R.da Rocha,
``From Dirac Action to\\ ELKO Action'',
\textit{Int.J.Mod.Phys.A}\textbf{24}, 3227 (2009).
\bibitem{daRocha:2008we}
R.da Rocha, J.M.Hoff da Silva, ``ELKO, flagpole and\\ flag-dipole spinor fields, and the instanton Hopf\\ fibration'', \textit{Adv. Appl. Clifford Algebras} \textbf{20}, 847 (2010).
\bibitem{daRocha:2013qhu}
R.da Rocha,L.Fabbri,J.M.Hoff da Silva,R.T.Cavalcanti, J.A.Silva-Neto,
``Flag-Dipole Spinor Fields in ESK Gravities'',
\textit{J.Math.Phys.}\textbf{54},102505(2013).
\bibitem{Fabbri:2016msm}
L.Fabbri, ``A generally-relativistic gauge\\ classification of the Dirac fields'',\\ \textit{Int.J.Geom.Meth.Mod.Phys.}\textbf{13},1650078(2016).
\bibitem{Fabbri:2018crr} 
L.Fabbri, ``Covariant inertial forces for spinors'',\\ 
\textit{Eur.Phys.J.C}\textbf{78}, 783 (2018).
\bibitem{Fabbri:2016laz}
L.Fabbri, ``Torsion Gravity for Dirac Fields'',\\ 
\textit{Int.J.Geom.Meth.Mod.Phys.}\textbf{14},1750037(2017).
\bibitem{Fabbri:2019kfr}
L.Fabbri, ``Polar solutions with tensorial connection of\\ the spinor equation'', \textit{Eur.Phys.J.C}\textbf{79}, 188 (2019).
\bibitem{Fabbri:2017pwp}
L.Fabbri, ``General Dynamics of Spinors'',\\ 
\textit{Adv. Appl. Clifford Algebras}\textbf{27}, 2901 (2017).
\bibitem{Fabbri:2020ypd}
L.Fabbri, ``Spinors in Polar Form'', arXiv:2003.10825
\bibitem{Lattanzi:2009mg}
M.Lattanzi, S.Mercuri, ``A solution of the strong CP\\
problem via the Peccei-Quinn mechanism through the\\
Nieh-Yan modified gravity and cosmological\\
implications'', \textit{Phys.Rev.D}\textbf{81},125015 (2010).
\bibitem{Castillo-Felisola:2015ema}
O.Castillo-Felisola, C.Corral, S.Kovalenko, I.Schmidt,\\ 
V.E.Lyubovitskij, ``Axions in gravity with torsion'',\\
\textit{Phys.Rev.D}\textbf{91}, 085017 (2015).
\bibitem{Fabbri:2020drl}
L.Fabbri, ``Re-normalizable Chern-Simons\\ Extension of Propagating Torsion Theory'',\\ \textit{Eur.Phys.J.Plus}\textbf{135}, 700 (2020).
\bibitem{Fabbri:2014dxa}
Luca Fabbri, ``A discussion on the most general torsion-gravity with electrodynamics for Dirac spinor matter\\ fields'', \textit{Int.J.Geom.Meth.Mod.Phys.}\textbf{12}, 1550099 (2015).
\end{thebibliography}
\end{document}